\documentclass[12pt]{article}

\usepackage[margin=0.7in]{geometry}

\usepackage[utf8]{inputenc}
\usepackage[T1]{fontenc}
\usepackage[english]{babel}
\usepackage{graphicx}
\usepackage{subcaption}

\usepackage{amsmath, amssymb, amsthm}

\usepackage{graphicx}
\usepackage{xcolor}

\usepackage[colorlinks=true,linkcolor=blue,citecolor=blue,urlcolor=blue]{hyperref}

\title{\boldmath Closing the dark photon window to thermal dark matter}
\author{
    Leon M. G. de la Vega\thanks{\textit{Instituto de F\'{\i}sica, Universidad Nacional Aut\'onoma de M\'exico, A.P. 20-364, Ciudad de M\'exico 01000, M\'exico.}} \thanks{\textit{Physics Division, National Center for Theoretical Sciences, Taipei 10617, Taiwan.}}  \and
    R.~Ferro-Hernandez\thanks{\textit{PRISMA+ Cluster of Excellence and  Institute for Nuclear Physics
Johannes Gutenberg University, 55099 Mainz, Germany.}} \and
    A.~Garc\'{\i}a-Viltres \footnotemark[1] \and
    Eduardo Peinado\footnotemark[1] \thanks{\textit{Departamento de F\'isica, Centro de Investigaci\'on y de Estudios Avanzados del Instituto Polit\'ecnico Nacional.}} \and
    E.~V\'azquez-J\'auregui\footnotemark[1] }

\begin{document}

\maketitle

\begin{abstract}
    
The nature of dark matter remains a central question in particle physics, cosmology, and astrophysics. The prevailing hypothesis postulates that dark matter consists of particles that interact only weakly with Standard Model particles. However, the knowledge of dark matter properties beyond these interactions is limited. This study explores a scenario involving a dark photon as a mediator between dark matter and the Standard Model, akin to the photon's role in electromagnetism. Recent cosmological and experimental evidence impose constraints on this scenario, focusing on results from direct detection experiments such as PICO-60, XENON-1T, and PANDAX-4T.
The results reveal severe constraints, effectively closing the window for laboratory searches for dark photons as mediators between the Standard Model and the dark sector (dark electrons) in the secluded dark matter scenario. The findings underscore the need for alternative explanations and offer fresh perspectives on the ongoing quest to understand dark matter and its interactions since they are nearly independent of the dark electron fraction content for the total dark matter. This analysis significantly narrows down the parameter space for thermal dark matter scenarios involving a dark photon portal, reinforcing the urgency of exploring alternative models and designing new experiments to unravel the mysteries surrounding the nature of dark matter.
\end{abstract}

\clearpage

A main open question in particle physics, cosmology, and astrophysics is the nature of dark matter (DM), which is a feebly interacting substance accounting for $\sim85\%$ of the matter in the Universe \cite{Planck:2018vyg}. A leading hypothesis is that DM is made of particles interacting very weakly with the Standard Model (SM) particles. Beyond this, very little is known about DM. Among the known pieces of information is that it does not move at relativistic speeds, hence the name "cold" or "warm" dark matter. It is frequently theorized that the DM particles were in thermal equilibrium with the SM particles in the early Universe. The dark particles decoupled from the thermal bath as the Universe expanded and cooled down. Furthermore, the abundance of DM particles after decoupling depends on their interactions and mass. Direct detection experiments have provided tight constraints on the interactions of DM with the SM particles, which severely constrain the allowed parameter space.

The search for physics beyond the Standard Model has been a long run task, with experiments following theoretical developments over the last few decades. A natural extension is the addition of a new gauge symmetry. This idea has attracted interest from the physics community, particularly as a possible solution to the mystery of DM. An extra Abelian gauge group is one of the simplest  extensions~\cite{Holdom:1985ag,Fayet:1980ad,Fayet:1980rr,Fayet:1990wx,Okun:1982xi,Georgi:1983sy}, which introduces a new gauge boson known as the dark photon, which couples to the SM through kinetic mixing. The dark photon itself constituting dark matter is a scenario that has been widely studied~\cite{Nelson:2011sf,Arias:2012az} and explored~\cite{PhysRevD.104.095029}. However, in the same way the photon mediates the electromagnetic interaction between electrons and protons, the dark photon could serve as the mediator of a new force, with the exciting prospect of dark matter being akin to a dark electron. 
In this letter, the scenario where the dark electron $\chi$ was in thermal equilibrium with the SM particles through the exchange of the dark photon $A'$, in the $m_\chi>m_{A'}$ regime, is severely constrained by the latest cosmological and experimental evidence, primarily results from the direct detection dark matter experiments PICO-60, XENON-1T, and PANDAX-4T. This finding nearly closes the window to laboratory searches for this scenario, highlighting the need for research into alternate explanations. Furthermore, these results bring a fresh perspective to a topic that continues to draw the attention of scientists and underscore the need for continued exploration into the nature of dark matter and its interactions.

In this scenario, the new gauge boson is the DM portal, which mixes kinetically with the $U(1)_Y$ gauge boson. The relevant Lagrangian is
\begin{equation}
{\cal L}=-\frac{1}{4}B_{\mu\nu}^\prime B^{\prime\mu\nu}-\frac{\epsilon}{2\cos\theta_W}B_{\mu\nu}^\prime B^{\mu\nu}  +\bar{\chi}(\gamma^\mu D_\mu+M_\chi) \chi +\mathrm{h.c}.
\label{lagrangian}
\end{equation}
where $M_\chi$ is the dark matter mass and the covariant derivative is $D_\mu=\partial_\mu+i Q_\chi g_D B^\prime_{\mu}$, $B^\prime_\mu$ is the new gauge boson, $g_D$ is the dark gauge coupling, $Q_\chi$ is the dark charge of the field $\chi$, and $B^{\mu \nu}$ is the strength tensor of the SM hypercharge gauge boson. Considering the case where $\chi$ is a Dirac fermion, like the electron, implies $Q_\chi\neq 1/2$. In this way, the dark fermion only interacts with the dark gauge boson and is stable, making it a potential candidate for DM.  The dark photon mass, $M_D$, can be generated by spontaneously breaking the $U(1)_D$ or through the Stueckelberg mechanism~\cite{ref1,HAN2020115154}. Here, it is considered that the physics mechanism giving rise to the dark photon mass lies on a higher scale than TeV and has no effect on dark matter direct detection or freeze-out.
The vector coupling of the dark photon to the Standard Model fermions, in the basis where the propagators are diagonal is \begin{equation}
g^{f}_V=e\epsilon \left[Q_f+\frac{1}{2\cos^2\theta_W}\frac{M^2_D}{M^2_Z-M^2_D}\left(t^{f}_3-2\sin^2\theta_W Q_f\right)\right], 
\end{equation}
where $e$ is the electron charge, $Q_f$ is the weak charge, $\theta_W$ is the Weinberg angle, and $M_Z$ is the $Z$ boson mass.
For a nucleus with an atomic number $Z$ and a mass number $A$, the coupling is \begin{equation}
g_V=e\epsilon Z \left[1+\frac{1}{2\cos^2\theta_W}\frac{M^2_D}{M^2_Z-M^2_D}\left(1-\frac{A}{2Z}-2\sin^2\theta_W \right)\right].
\end{equation}
When $M_D\ll M_Z$, the usual dark photon scenario is recovered, namely $g_V\approx e\epsilon Z$. For a xenon target, $Z=54$ and $A=124-136$, hence, the interaction becomes stronger for $M_D>M_Z$ and weaker for $M_D<M_Z$. In terms of constraints on the kinetic mixing term $\epsilon$, this implies that for $M_Z< M_D$, the contribution of the $Z$ boson vector coupling improves the limits on $\epsilon$ by $\sim 50\%$ with respect to only taking into account $e\epsilon Z$. A similar effect occurs for the target used in PICO-60, either carbon or fluorine in the  C$_{3}$F$_{8}$ target material.

\begin{figure}[htpb!]
  \centering
  \includegraphics[scale=0.35]{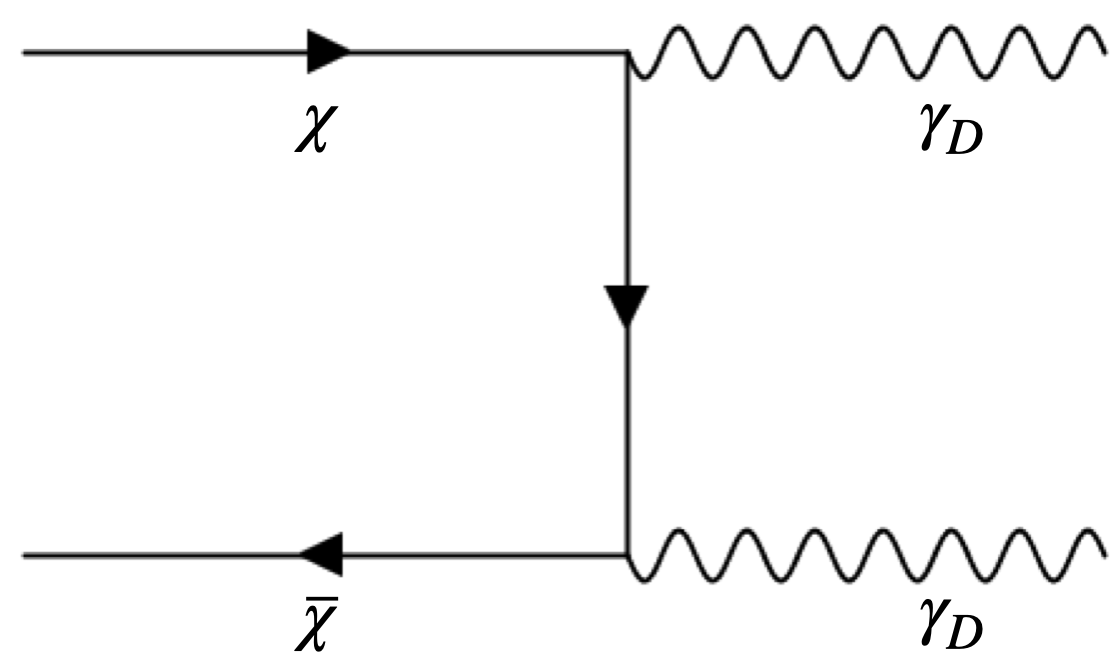}
    \caption{Tree level Feynman diagram of dark matter annihilation in the Early Universe.}
    \label{fig:tchannelannihilation}
\end{figure}

When the DM candidate is heavier than the dark photon, the dominant annihilation channel of dark matter in the early Universe is $\chi \chi \rightarrow \gamma_D \gamma_D$ \cite{Pospelov:2007mp,Pospelov:2008zw}. This annihilation can lead to the freezing-out of DM at the observed dark matter relic density. The scattering amplitude is obtained at the tree level from the Feynman diagram in Fig. \ref{fig:tchannelannihilation}.

\begin{figure}[htpb!]
\centering
    \includegraphics[scale=0.75]{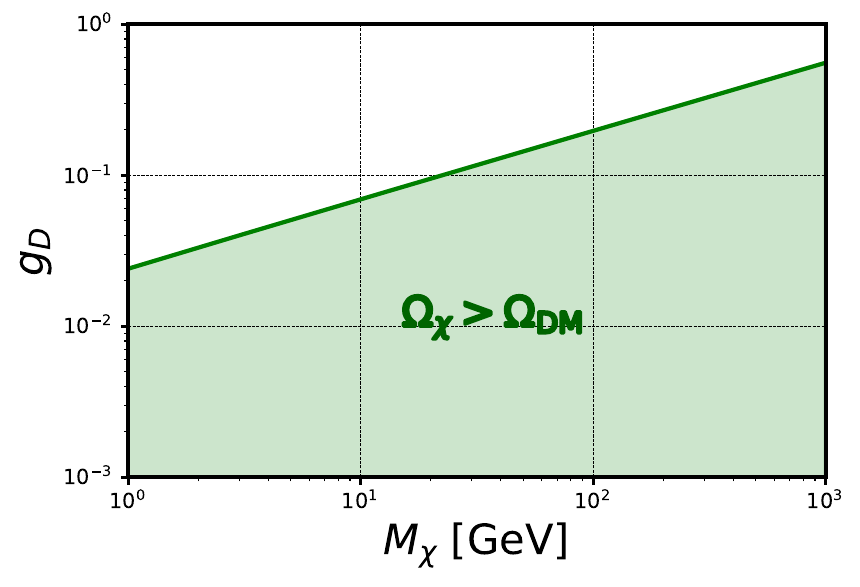}
    \caption{Numerical relationship between $M_\chi$ and $g_D$. The green line corresponds to the observed value of the dark matter relic density.}
    \label{fig:numerical}
\end{figure}

The annihilation cross-section relevant for the thermal freeze-out of DM is \cite{Pospelov:2007mp}
\begin{equation}
    \sigma v =\frac{g_D^4}{16 \pi M_\chi^2}\sqrt{1-\frac{M_{D}^2}{M_\chi^2}}.
\end{equation}
The resulting DM relic density is insensitive to the dark photon mass unless the masses of both are finely tuned to a close value. This allows to determine a relationship between $g_D$ and $M_\chi$ permitted by the observed DM relic density, subject to the conditions $M_\chi>M_D$ and $M_D^2/M_\chi^2\nsim 1$. Using micrOMEGAs~\cite{Belanger:2018ccd}, a numerical estimate of this relationship~\cite{delaVega:2022uko} was obtained, and crosschecked with the analytical approximation of the secluded WIMP scenario \cite{Pospelov:2007mp}. The numerical solution, shown in Fig. \ref{fig:numerical}, can be expressed as \begin{equation}
g_D(M_{\chi})\approx \frac{1}{40}\left(\frac{M_\chi}{\mathrm{GeV}}\right)^{0.45}-0.00822. \label{eq:gdrelic}
\end{equation}

In the region of interest for DM mass 1 GeV $<M_\chi<$ 1000 GeV, the dark gauge coupling is in the range 0.01 $<g_D<$ 1. The freeze-out annihilation channel of dark matter produces dark photons, which could constitute a relativistic degree of freedom in the early Universe. The amount of relativistic degrees of freedom in the radiation-dominated Universe control the rate of expansion, and Big Bang Nucleosynthesis (BBN) is highly sensitive to them. To avoid contradicting BBN measurements of $N_{eff}$ (the effective number of relativistic species), a dark photon mass larger than $10$ MeV and a lifetime smaller than $\mathcal{O}(1s)$ is imposed, which sets a lower bound on $\varepsilon$. A lower limit constraint is obtained from the requirement of thermal equilibrium for dark photons decaying to electrons, which thermalizes the dark sector in the early Universe, leading to the freeze-out scenario assumed in this work. Requiring that the decay rate is larger than the Hubble rate at freeze-out leads to \cite{Pospelov:2007mp}

\begin{equation}
    |\varepsilon| \gtrsim 10^{-6}\sqrt{\left(\frac{10\text{GeV}}{M_D}\right)}\left(\frac{M_\chi}{500\text{GeV}}\right).
\end{equation}
For $M_\chi=316$ GeV, $M_D=10^{-3}$ GeV, a constraint of $|\varepsilon|\gtrsim 10^{-4}$ is obtained, and for larger dark photon masses, this constraint is relaxed. 

Different experiments constrain the kinetic mixing depending on the dark photon's mass. For a dark photon heavier than $\mathcal{O}(\text{MeV})$ but lighter than $\mathcal{O}(\text{GeV})$, the strongest experimental constraints come from beam dump experiments \cite{Bross:1989mp, Riordan:1987aw, Bjorken:1988as, Batell:2014mga, Marsicano:2018krp, Blumlein:2011mv, Blumlein:2013cua, Gninenko:2012eq} and cosmological observations \cite{Fradette:2014sza}. On the other hand, for higher masses, collider experiments \cite{LHCb:2019vmc, CMS:2019kiy, BaBar:2014zli, Curtin:2014cca, BaBar:2017tiz} provide the best constraints. These limits depend on the branching ratio of the dark photon to visible and invisible decays. In the secluded scenario presented here, the constraints will mainly come from dark photon decays to visible particles.

Kinetic mixing can also be accessed with direct detection experiments through coupling to the protons and electrons in the target material. These experiments can impose bounds complementary to beam dump and collider experiments, as well as cosmological observations.\\

Direct detection experiments measure nuclear recoils from nuclei-DM scattering. These experiments require an incredibly low background environment and are installed in underground laboratories. The detectors measure tiny energy deposits, which may indicate the interaction with dark matter particles. Highly sensitive detectors such as cryogenic bolometers, noble-liquid, or super-heated detectors are used to detect these small energy deposits. The increasing sensitivity of the detectors and the continued search for new detection methods allow for reaching extremely low cross sections, below $10^{-47}$ cm$^2$ for dark matter masses of 20-30 GeV. Currently, no convincing signal of dark matter has been observed. However, the negative results further constrain the strength of the DM-SM interaction.

Traditionally, two couplings have been explored. Most of the experiments nowadays are mostly sensitive to a spin-independent coupling between the nucleons and the dark matter where the cross-section is proportional to the number of nucleons squared in the target material~\cite{BERTONE2005279,LEWIN199687,RevModPhys.85.1561,PhysRevD.69.063503,Marcos_2016,PhysRevLett.95.101301}. These couplings have been widely probed in mass ranges from a few GeV up to the TeV scale. The searches for this interaction are led by xenon Time Projection Chambers (TPC), namely, PANDAX-4T~\cite{PhysRevLett.127.261802}, XENON-1T~\cite{PhysRevLett.121.111302}, and LZ~\cite{PhysRevLett.131.041002}. These detectors are a dual-phase TPC filled with xenon. The target material is inside a stainless steel or titanium vessel in a cylindrical shape. Two arrays of photomultiplier tubes are positioned at the ends of the cylindrical vessel to register light signals produced by the events. An event consists of flash of scintillation light (prompt signal), followed by a flash of electroluminescence (delayed signal). The detectors are immersed in active and passive shielding vetoes consisting of water and gadolinium loaded scintillator.

In addition, searches have also focused on couplings proportional to the spin of the nucleus. The cross-section for this spin-dependent interaction is proportional to the spin of an unpaired nucleon~\cite{BERTONE2005279,LEWIN199687,RevModPhys.85.1561,PhysRevD.69.063503,Marcos_2016,PhysRevLett.95.101301}. The PICO bubble chambers~\cite{PhysRevD.100.022001,PhysRevLett.118.251301} are the leading technology for couplings to protons employing fluorocarbon materials as targets. These detectors employ flurocarbon materials (CF$_{3}$I or C$_{3}$F$_{8}$) and consist of a high purity vessel made of fused silica filled with the target material. The system is inside a stainless steel pressure vessel filled with an hydraulic fluid. The detector is located inside a water tank to provide shielding from external backgrounds, also acting as a thermal bath. The data acquisition system consists of cameras to photograph the detector and acoustic sensors (piezoelectric transducers) to register the bubble formation process. These detectors are insensitive to electron recoils by adjusting the thermodynamic parameters (pressure and temperature), while alpha decays are discriminated against nuclear recoils using the acoustic parameter.

All experimental results have been consistent with no dark matter signal and constraints have been established for the interaction cross-section between nucleons and dark matter~\cite{Billard_2022,akerib2022snowmass2021,adams2023axion,carney2023snowmass2021,MITRIDATE2023101221}. 

The latest results from the leading spin-dependent and spin-independent results, namely, PICO-60~\cite{PhysRevD.100.022001,PhysRevLett.118.251301}, XENON-1T~\cite{PhysRevLett.121.111302}, and PANDAX-4T~\cite{PhysRevLett.127.261802}, can be reinterpreted under the scenario discussed in this letter. The relevant cross-section employed takes the form 
\begin{equation}
\frac{d\sigma_T}{dE_R}=\frac{M_N}{2\pi v^2}\frac{g^2_D g^2_V F^2_{SI}(E_R)}{\left(q^2+M^2_D\right)^2},
\end{equation}
where $M_N$ is the mass of the target,  $F_{SI}$ is a form factor that considers that the nucleus is spatially extended, and $E_R$ is the recoil energy. 

\begin{figure}[htpb!]
\centering
    \includegraphics[scale=0.25]{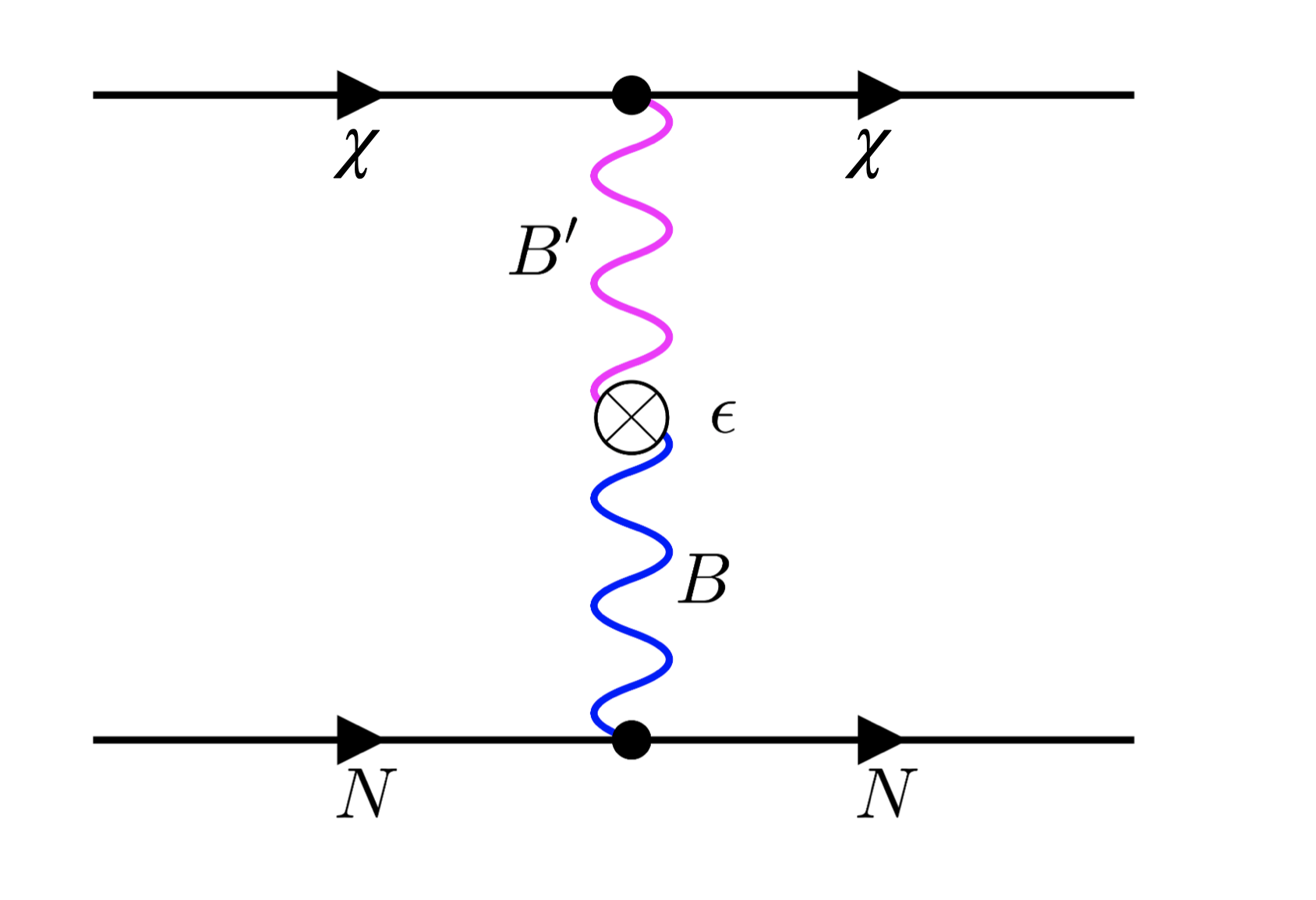}
    \caption{DM direct detection interaction with a
    dark photon coupled to the SM through kinetic mixing.}
    \label{fig:diagramDD}
\end{figure}

Dark matter direct detection experiments set constraints on the product $g_D \epsilon$, as seen from the Feynman diagram in Fig. \ref{fig:diagramDD}. Therefore, given that the dark matter relic density fixes $g_D$, these constraints are directly translated into a bound on $\epsilon$.

\begin{center}
\begin{figure}
\begin{subfigure}{0.5\textwidth}
\centering
\includegraphics[width=\textwidth]{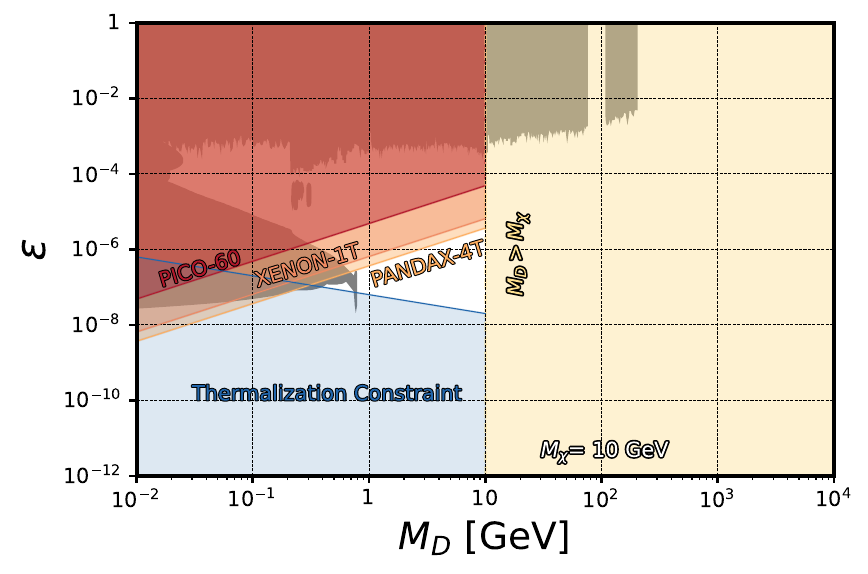}
\caption{}
\label{fig:sub1}
\end{subfigure}%
\begin{subfigure}{0.5\textwidth}
\centering
\includegraphics[width=\textwidth]{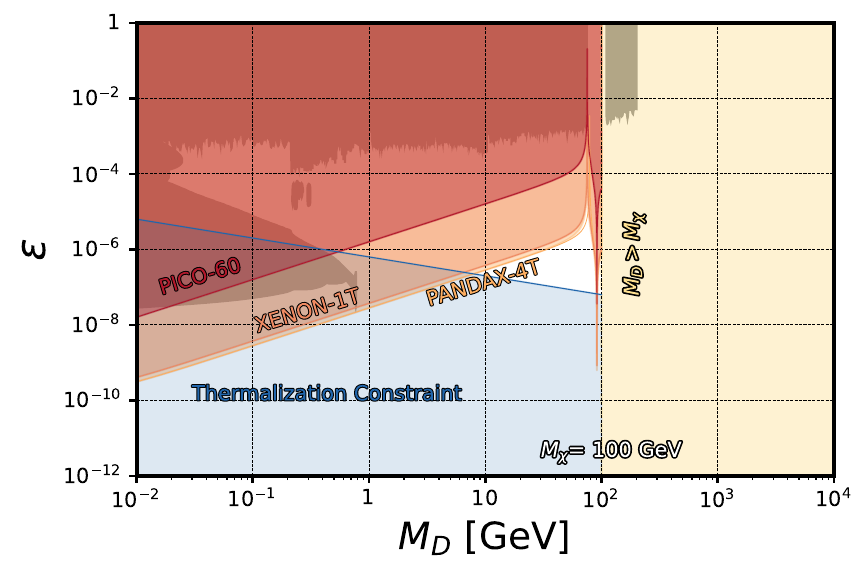}
\caption{}
\label{fig:sub2}
\end{subfigure}\\[1ex]
\centering
\begin{subfigure}{0.5\textwidth}
\includegraphics[width=\textwidth]{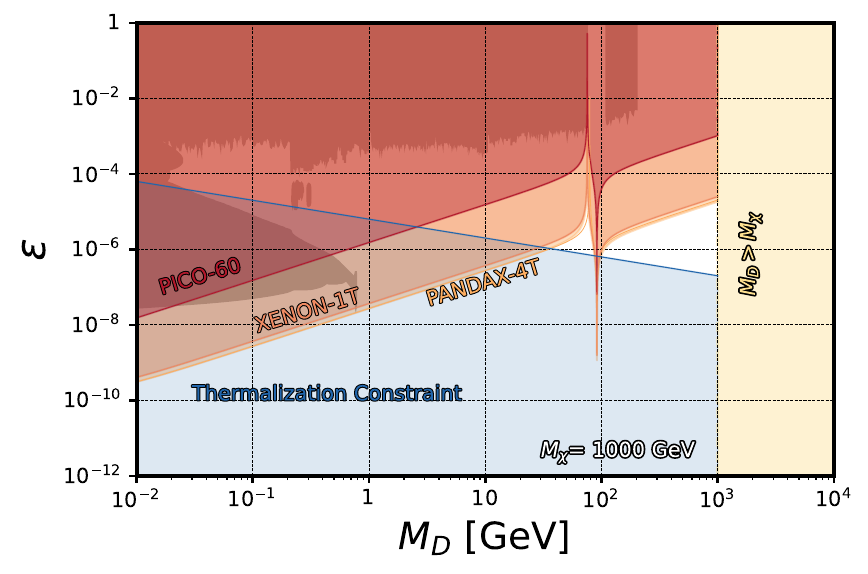}
\caption{}
\label{fig:sub3}
\end{subfigure}
\caption{Constraints on the kinetic mixing parameter $\epsilon$ as function of the dark photon mass for three representative values of the dark matter mass $M_\chi = 10$, $100$, and $1000$ GeV.
}
\label{fig:epsconstraints}
\end{figure}
\end{center}

Fig. \ref{fig:epsconstraints} illustrates the constraint on $\epsilon$ as function of the dark photon mass for three representative values of $M_\chi$. There is almost a complete exclusion of the parameter space for the cases presented. The secluded scenario constrains the yellow region, which requires $M_\chi>M_D$. The blue region sets a lower limit on the size of $\epsilon$. It is derived from the thermalization constraint, namely that it must interact sufficiently with the Standard Model particles to yield thermal dark matter. Lastly, the red region is constrained by direct DM detection experiments. 

\begin{figure}
\begin{subfigure}{0.5\textwidth}
\centering
\includegraphics[width=\textwidth]{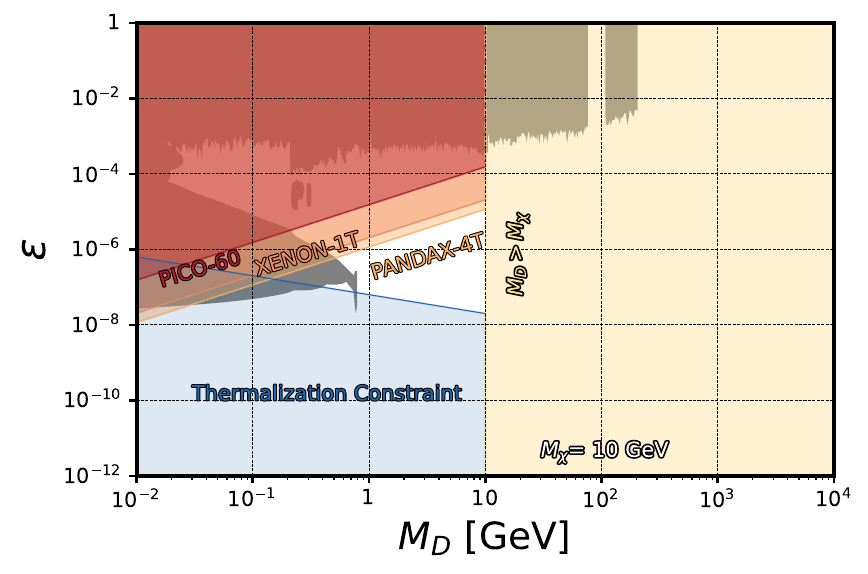}
\caption{}
\label{fig:onepercent1}
\end{subfigure}%
\begin{subfigure}{0.5\textwidth}
\centering
\includegraphics[width=\textwidth]{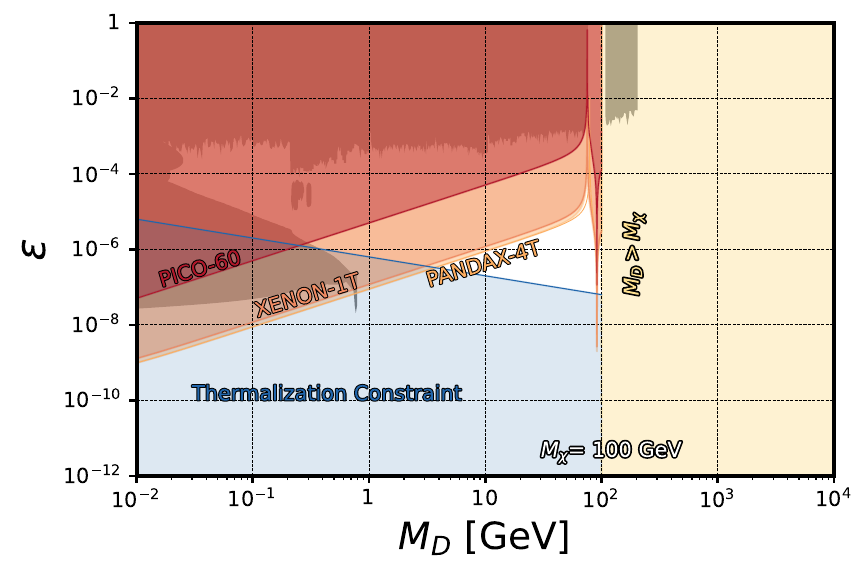}
\caption{}
\label{fig:onepercent2}
\end{subfigure}\\[1ex]
\centering
\begin{subfigure}{0.5\textwidth}
\includegraphics[width=\textwidth]{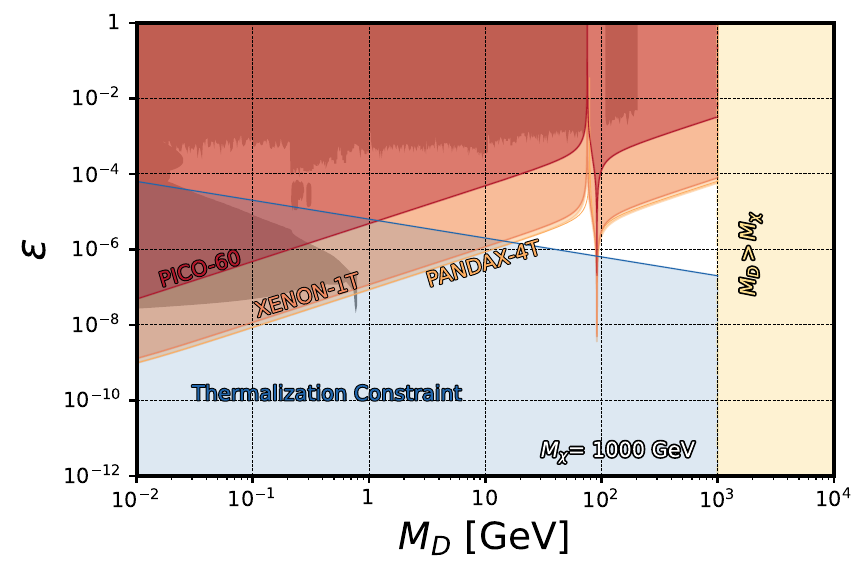}
\caption{}
\label{fig:onepercent3}
\end{subfigure}

\caption{Constraints on the kinetic mixing parameter $\epsilon$ as function of the dark photon mass for three representative values of the dark matter mass $M_\chi = 10$, $100$, and $1000$ GeV. The dark electron is assumed to account for only 1\% of the dark matter relic density in this case.}
\label{fig:onepercent}
\end{figure}

Furthermore, the constraints remain nearly unchanged if the dark electron accounts for a small fraction of the DM in the Universe. For example, Fig. \ref{fig:onepercent} illustrates the three representative cases studied for a fraction of dark electrons of $1\%$.

\section*{Conclusions}

Direct detection experiments combined with cosmological constraints significantly reduce the parameter space of the dark photon. These constraints are even stronger than current and future dark photon searches, meaning that the window for the secluded dark photon portal to thermal dark matter is getting closed. These constraints remain nearly unchanged, even when the dark electron constitutes only a small fraction of the total dark matter in the Universe. This arises from the partial cancellation between two effects. The first is the need of an efficient annihilation process to achieve a lower relic density, which increases $g_D$,  enhancing the efficiency of direct dark matter detection experiments. The second is that a smaller fraction implies a smaller local density of dark electrons, which in turn diminish the sensitivity of direct detection experiments. Even in the extreme scenario where the local dark matter density is overestimated by a factor of two, the constraints imposed by direct detection experiments still almost completely close the allowed parameter space. Future low-energy experiments aiming to probe the existence of a light-dark photon will observe no results if this dark photon is the portal between a thermal relic dark electron and the Standard Model. Additionally, future direct detection experiments, such as ARGO and DARWIN, will explore the remaining available parameter space, having the potential to close off the secluded dark photon portal to thermal dark matter.

\section*{Acknowledgements}
\indent The authors would like to thank the PICO collaboration and Jens Erler for useful comments and discussions. This work is supported by the German-Mexican research collaboration grant SP 778/4-1 (DFG) and 278017 (CONACYT), the projects CONACYT CB-2017-2018/A1-S-13051 and CB-2017-2018/A1-S-8960,  DGAPA UNAM grants PAPIIT-IN107621, PAPIIT-IN107118, PAPIIT-IN108020, and PAPIIT-IN105923, Fundación Marcos Moshinsky, the National Science and Technology Council, the Ministry of Education (Higher Education Sprout Project NTU-112L104022) and the National Center for Theoretical Sciences of Taiwan. E.P and E.V.J. are grateful for the support of PASPA-DGAPA, UNAM for a sabbatical leave. RF thanks Instituto de F\'{\i}sica UNAM for the hospitality during his visit\\

\bibliographystyle{unsrt}
\bibliography{bibliography.bib}

\end{document}